\documentclass[12pt]{JHEP3}
\newcommand{\COMMENTO}[1]{}
\def \f {{\rm f}}
\def\x'{\mathaccent 19 x}
\def\th'{\mathaccent 19 \theta}
\def\barth'{\mathaccent 19 {\bar{\theta}}}

\def\ipr{{i^\prime}}

\def\ket#1{|#1\rangle}
\def\bra#1{\langle#1|}
\def\sgn{\mathrm{sgn}}

\def\ad{{\dot a}}
\def\bd{{\dot b}}
\def\cd{{\dot c}}

\preprint{DFTT xx/02}
\preprint{\hfill February 2002}

\title{
Boundary States for GS superstrings
in an $Hpp$ wave background.
\footnote{Work supported in
part by the European Community's Human Potential Programme under
contract  HPRN-CT-2000-00131 Quantum Spacetime.}
}

\author{
Marco Bill\'o \footnote{e-mail:billo@to.infn.it},
Igor Pesando\footnote{e-mail:ipesando@to.infn.it}
\\
Dipartimento di Fisica Teorica, Universit\`a di Torino\\
 Via P.Giuria 1, I-10125 Torino, Italy\\
 and I.N.F.N., Sezione di Torino
}

\abstract{
We construct the boundary states preserving half the global
 supersymmetries in string theory propagating on a
$Hpp$ background.
}

\begin{document}

\section{Introduction and conclusions.}
Soon after the conjecture of  AdS/CFT correspondence \cite{Maldacena:1997re}, 
some authors addressed the problem of solving the
string theory in $AdS\times S$ backgrounds.
The first approach \cite{strings} made use of the so-called ``killing gauge''
for $\kappa$-symmetry in order to get the string action in
these backgrounds. The actions turned out to be
almost intractable, so a renewed attempt was made, using this time
light cone gauge \cite{stringsLC}; the models still remained not solvable.
However, it was recently pointed out \cite{rrm} that
a  conformal  model describing type IIB  superstring
propagating  in a particular  wave  metric supported by
a  {\it Ramond-Ramond}   5-form background \cite{bla}:
\begin{eqnarray}
\label{bem}
 ds^2 = 2 dx^+dx^- &&- \f^2 x_I^2 dx^+ dx^+  + dx^I dx^I \
,   \ \ \ \ \ \ \  I=1,..., 8 \ ,
\\
\label{bemm}
&&F_{+1234}= F_{+5678}= 2\f
\end{eqnarray}
is exactly solvable.
This  background  has several remarkable properties.
It preserves the maximal number of 32 supersymmetries \cite{bla},
and it is related by a Penrose limit
to the $AdS_5 \times S^5$  background \cite{bla,Hatsuda:2002xp}.
It is worth noticing that  the metric global symmetry $O(8)$ is broken
to $SO(4)\times SO(4)$ by the RR background.

The same properties are true for other backgrounds \cite{Russo:2002rq},
\cite{Berenstein:2002jq}
and we therefore expect that our analysis extends to these cases.

Soon after the discovery of the exact solvability of this string model,
an interpretation of its  states
within the AdS/CFT correspondence \cite{Berenstein:2002jq} as states
with large angular momentum $J\sim \sqrt N$ and $J-\Delta$ finite in the dual
${\cal N}=4$ SYM. Notice however that the string Hamiltonian in this background
$H_{\mbox{lc string}}$ explains only
the leading order anomalous dimension:
$\Delta=H_{\mbox{lc string}}+O\left( 1/R \right)$.
This interpretation has been
extended to other conformal cases in \cite{otherCFTAdS}.

In this article we derive the boundary states associated with branes
which preserve 16 supercharges in this solvable background.
Differently from what happens in the flat background, where both the
covariant NSR formulation (see \cite{Billo:1998vr} for a complete and
consistent set of normalizations) and light-cone  GS formulation
\cite{Green:1996um} are available, in this case only the latter
is at our disposal.

The plan of the paper is the following. In section 2 we give an
heuristic derivation of the $\kappa$ gauge fixed action found by
Metsaev and we discuss why it is possible to fix the light-cone gauge.
In section 3 after discussing the allowed boundary conditions on the
bosonic fields $x^I$ we construct the bosonic boundary states. They
turn out to be $O(8)$ invariant, just as the metric is.
Finally in section 4 we investigate the conditions to be imposed in the
fermionic sector
in order to break exactly half the number of global supersymmetric charges.
These conditions break explicitly the global
symmetry down to $SO(4)\times SO(4)$, as it could be expected due to
``fermionic'' nature of the RR background.
It turns out that $D(-1)$ and $D7$ break more than 16 supersymmetries
(actually 24)
and that the other branes of type $IIB$ superstring can only have
special embeddings and
must seat at the origin of the transverse coordinates 
if they must
preserve 16 global supersymmetry charges. They can instead sit anywhere if 
they are left invariant by 8 charges only.

The branes that can be described on the light cone in this
background are always ``instantonic'' and parallel to the wave.
Since the generators $J^{-I}$ do not correspond to isometries, these branes
are not in an obvious way related to the ones perpendicular to the wave which
are difficult, if not impossible, to describe in the light-cone formalism
because even if we could reach the gauge $x^-=p^-\tau$ the resulting
$\sigma$ model would still be interacting.
In a similar way it is difficult to perform a double Wick rotation
\cite{Morales:1997hk} to obtain ``physical'' branes.
It would hence be desirable to solve this theory in other gauges or in
a pure spinor formalism \cite{Berkovits:2001ue} since the solution
in NSR formalism appears to be difficult.
A better understanding of why the D-instanton seems to break more that
16 charges would also be desirable, in particular to clarify whether this
happens only due to our ansatz or it because of a more fundamental reason.

Another interesting point would be to compute the $1/R$ corrections to
the GS light-cone action in order to find the corrections to the
anomalous dimensions in the dual SYM.

\section{Green-Schwarz superstrings in an $Hpp$-wave.}
We start considering the lagrangian for a bosonic string propagating
on the $Hpp$ wave background (\ref{bem}) in the conformal gauge:
\begin{equation}
\label{L_bos}
{\cal L}=\frac{1}{2}
\left(
2\partial_A x^+ \partial^A x^-
-\f^2 x_I^2 \partial_A x^+ \partial^A x^+
+ \partial_A x^I \partial^A x^I~.
\right)
\end{equation}
The variation with respect to $x^-$ of the action eq. (\ref{L_bos}) implies
that $x^+$ obeys the massless free scalar equation; we can therefore impose the
light cone gauge (we fix conventionally $2\alpha'=1$)
\begin{equation}
\label{lc_gauge}
x^+=p^+ \tau~.
\end{equation}
In this gauge the previous lagrangian becomes, setting $m =p^+\f$,
\begin{equation}
\label{L_bos_gf}
{\cal L}=\frac{1}{2}
\left(\partial_A x^I \partial^A x^I
-m^2  x_I^2
\right),
~.
\end{equation}
Furthermore, we must take into account the Virasoro conditions, giving
in this gauge the following two constraints:
\begin{eqnarray}
\label{x-s} \x'^-&=&
\frac{\dot x^I\, \x'^I}{p^+}~,
\\
\label{x-t}
\dot x^-&=&
-\frac{\dot x^{I2}+ \x'^{I2}-m^2 x^{I2}}{2p^+}~.
\end{eqnarray}

The gauge fixed lagrangian (\ref{L_bos_gf}) is that of 8 massive
two-dimensional free bosons. In view of the fact that the background preserves
32 real supersymmetric charges and that it reduces to flat space in the
limit $m\rightarrow0$ it is  natural to expect that the
fermionic lagrangian (which was first derived in \cite{rrm} via a
super-coset construction) corresponds to the modification of
the usual flat-space GS action by mass terms. Since the
the RR field in the background eq. (\ref{bemm})  breaks the
$SO(8)$ symmetry to $SO(4)\times SO(4)$, the mass term
contains the product
$  \Pi
\equiv
{\gamma}^1 \bar{\gamma}^2 {\gamma}^3 \bar{\gamma}^4$,
which is a symmetric matrix so that the mass term must be
``mixed''. The fermionic lagrangian reads thus
 \begin{equation}
\label{L_fer_0}
{\cal L}_{F} =
\frac{i}{2}\left(
S^a\partial_+ S^a
+{\tilde S^a} \partial_- {\tilde S^a}
\right)
-{\rm i}m\,S^a  \Pi_{ab} {\tilde S}^b~,
\end{equation}
where
$S^a$ and $\tilde S^a$ \footnote{
These fields are related to the ones used in
\cite{rrm} by $$\Gamma^{+-}\theta^1=\frac{1}{2^{3/4}}S~,
\,\,\,\,\,\,
\Gamma^{+-}\theta^2=\frac{1}{2^{3/4}}{\tilde S}
$$
}
($a=1,\ldots 8$) are the canonical GS fields and
$\partial_\pm =\partial_0 \pm \partial_1 =\partial_\tau\pm\partial_\sigma$.
The coefficient of the mass term was chosen in such a way that $x$ and
$S$ have the same plane wave expansion.

\section{The bosonic sector.}
\subsection{Boundary conditions and quantization in the bosonic sector.}
We start discussing the boundary conditions allowed by the
bosonic lagrangian (\ref{L_bos}); in particular the $x^I$ variation
implies
\begin{equation}
\delta x^I\, \x'^I|^{\sigma=1}_{\sigma=0}=0
\end{equation}
which allows all usual types of boundary conditions: periodic, Neumann and
Dirichlet.
The other variations imply
\begin{equation}
\delta x^-\, \x'^+|^1_0=
\delta x^+~,
\hskip 0.4cm
\left(\x'^- -  \f^2 x_I^2 \x'^+\right)|^1_0=0
\end{equation}
which are trivially satisfied when (\ref{lc_gauge}) and (\ref{x-s})
are used.
Moreover the condition (\ref{x-s}) implies that $x^-$
has always Dirichlet boundary conditions $\x'^-\ket B = 0$ at a boundary
inserted at fixed $\tau$ in the closed channel, both for Neumann and Dirichlet
conditions on the $x^I$, as it is true also for $x^+$ \cite{Green:1996um}.

The mode expansions for  the transverse coordinates $x_I$ corresponding to the
different boundary conditions are
\begin{eqnarray}
\label{x_cl}
x_{\rm cl}^I(\sigma,\tau) &=&
\cos(m) \tau\, x_0^I +m^{-1} \sin(m\tau)\, p_0^I
\nonumber\\
&&+ {\rm i}\sum_{n=\!\!\!\!/\,\, 0}
\frac{\sgn(n)}{\sqrt {|\omega_{n}|} }
\left( e^{-{\rm i}(\omega_n \tau -k_n\sigma)} a_n^{I}
 + e^{-{\rm i}(\omega_{n} \tau +k_n\sigma)} {\tilde a}_n^{I}
\right)~,
\\
\label{x_N}
x_{NN}^I(\sigma,\tau) &=&
\cos (m \tau)\, x_0^I +m^{-1} \sin (m\tau)\, p_0^I
\nonumber\\
&&+ {\rm i}\sum_{n=\!\!\!\!/\,\, 0}
\frac{\sgn(n)}{\sqrt {|\omega_{n(o)}|} }
e^{-i \omega_{n(o)} \tau}\, \cos\frac{k_n\sigma}{2} a_n^{I}~,
\\
\label{x_D}
x_{DD}^I(\sigma,\tau) &=&
\frac{-q_0^I\, \sinh(m(\sigma-1)) +q_1^I\, \sinh(m\sigma)}{\sinh(m)}
\nonumber\\
&&+{\rm i}\sum_{n\not= 0}
\frac{1}{\sqrt {|\omega_{n(o)}|} }
e^{-i \omega_{n(o)} \tau}\, \sin\frac{k_n\sigma}{2} a_n^{I}~,
\end{eqnarray}
where we have defined
\begin{equation}
\omega_{\pm |n|}=\pm \sqrt{k_n^2 + m^2 },
\,\,\,
\omega_{\pm |n|(o)}=\pm \sqrt{(k_n/2)^2 + m^2 },
\,\,\,
 k_n \equiv 2\pi n
\end{equation}
and we have fixed the Dirichlet b.c.'s as
$x_{DD}(\sigma=0)=q_0$  and $x_{DD}(\sigma=1)=q_1$.

The previous eq.s are normalized in such a way that
the usual commutation relation
\begin{equation}
[x^I(\sigma),{\cal P}^J(\sigma')]=i\, \delta(\sigma-\sigma')\, \delta^{IJ}~,
\end{equation}
where ${\cal P}^I=\dot x^I$, imply the following commutators for the modes:
\begin{equation}
[p_0^I,x_0^J]=-i\, \delta^{IJ}
\,,\qquad
[a_m^I,a_n^J] = [{\tilde a}_m^I,{\tilde a}_n^J]
= \sgn(n)\,\delta_{m+n,0}\, \delta^{IJ}~.
\end{equation}

The vacuum state of the bosonic part is the direct product of a  zero mode
vacuum and the Fock vacuum for string oscillation modes, and is altogether
defined by
\begin{equation}
\label{vacdef}
{a}_0^I|0\rangle=0\,,
\qquad
{a}_n^I|0\rangle=
\tilde{a}_n^I|0\rangle=0
\ , \ \ \ \ n=1,2,... \ ,
\end{equation}
where we introduced the zero mode creation and annihilation operators
\begin{equation}
a_0^{I\dagger} = \frac{1}{\sqrt{2 m}}(p_0^I + {\rm i}m x_0^I)
\,,\qquad
{a}_0^I =\frac{1}{\sqrt{2m}}( p_0^I - {\rm i}m x_0^I)~,
\end{equation}
as usual for an harmonic oscillator.

\subsection{Bosonic boundary states.}
\paragraph{Neumann  conditions.}
The Neumann boundary condition
$\left.\frac{\partial}{\partial\sigma_{open}}x \right|_{\sigma=0}=0$
for an open string field can be reinterpreted in the closed channel
as a condition on the state $\ket B$ that represents the insertion of the
boundary at a fixed $\tau$:
\begin{equation}
\dot x|_{\tau=0} \, \ket B=0~.
\end{equation}
This equation can be written in terms of the modes:
\begin{equation}
p_0\ket B= (a_n+{\tilde a}_{-n}) \ket B=0~,
\end{equation}
and can be satisfied by the following boundary state:
\begin{equation}
\label{B_N_bos}
\ket B=e^{-\frac{1}{2}\, a_0^{2\dagger}}\,
    e^{-\sum_{n=1}^\infty a_n^\dagger \tilde a_n^\dagger}
    \ket 0~,
\end{equation}
where the zero mode part has become structurally analogous to the non zero
mode part because the zero modes have acquired mass.

\paragraph{Dirichlet conditions.}
The Dirichlet  boundary condition in the closed channel
\begin{equation}
\label{D_bc}
\left.\left(x-q_0\right)\right|_{\tau=0}\ket B=0~,
\end{equation}
which also implies that $\x'|_{\tau=0} \, \ket B=0$,
becomes, in terms of the modes,
\begin{equation}
(x_0-q_0)\ket B= (a_n-{\tilde a}_{-n}) \ket B=0~.
\end{equation}
%
The solution to these equations is given by
\begin{equation}
\label{B_D_bos}
\ket B=e^{\frac{1}{2}\, (a_0^{\dagger}-i\sqrt{2m} q_0 )^2}\,
    e^{+\sum_{n=1}^\infty a_n^\dagger \tilde a_n^\dagger}
    \ket 0~,
\end{equation}
where again the zero mode part is structurally analogous to the non zero
mode part.
\paragraph{General situation.}
We can summarize the previous discussion by introducing a matrix
$M_{IJ}$ $={\rm diag}(\pm 1,\dots,\pm 1)$, every $-1$ ($+1$) entry 
corresponding
to a Neumann (Dirichlet) boundary condition on a transverse field:
\begin{equation}
\left(\partial_+ x^I - M_{IJ}\partial_- x^J\right)\ket B=0~.
\end{equation}
This equation can be  expressed on the operators as
\begin{equation}
\label{ans-a}
a^I_n - M_{IJ} {\tilde a}^J_{-n} \ket B=0
\end{equation}
and it is still valid for $n=0$ with the proviso that in the Dirichlet
case we choose $q_0=0$.
The corresponding boundary state (with $q^I_0=0$ for simplicity)
reads
\begin{equation}
\ket B=
  e^{\frac{1}{2}\, M_{IJ}\,a_0^{I\dagger}a_0^{J\dagger} }
  e^{+\sum_{n=1}^\infty\, M_{IJ}\,a_n^{I\dagger} \tilde a_n^{J\dagger}}
    \ket 0~.
\end{equation}
This description can obviously be generalized to any matrix $M_{(v)}=M^{IJ}$
belonging to $O(8)$.

\section{The complete theory.}
\subsection{Boundary conditions and quantization in the fermionic sector.}
The fermionic equations of motion which can be derived from
(\ref{L_fer_0}):
\begin{equation}
\partial_+ S  - m \Pi {\tilde S} =0
\,,\qquad
\partial_- {\tilde S}  +m \Pi S =0
\end{equation}
must be supplemented with boundary conditions satisfying the
constraint
\begin{equation}
\left.\left(
S^a\, \delta S^a
-
{\tilde S}^a\, \delta {\tilde S}^a
\right)\right|^{\sigma=1}_{\sigma=0}
=0~.
\end{equation}
The allowed boundary conditions are then the usual: periodic (i.e., closed
string) or generalized open, i.e. ${\tilde S}^a|_{\sigma=0}=S^a|_{\sigma=0}$ and
${\tilde S}^a|_{\sigma=1}=R^a_b\, S^b|_{\sigma=1}$, with  $RR^T=1$.
The mode expansions with closed string conditions (which are the ones relevant
for the boundary state construction) are
\begin{eqnarray}
 S^a(\sigma,\tau)
&=&
\cos(m\tau)\, S_0^a +\sin(m\tau)\  (\Pi {\tilde S}_0)^a
\nonumber\\
&&+ \sum_{n=\!\!\!\!/\,\, 0} c_n
\Bigl(
e^{-{\rm i}(\omega_n \tau -k_n\sigma)} S^a_n
+ {\rm i}{\textstyle{\omega_n-k_n\over m}}
e^{-{\rm i}(\omega_{n} \tau +k_n\sigma)}
\Pi {\tilde S}_n^a
\Bigr)~,
\\
 {\tilde S}^a(\sigma,\tau)
&=&
\cos(m\tau)\, {\tilde S}_0^a -\sin(m\tau)\  (\Pi { S}_0)^a
\nonumber\\
&&+ \sum_{n=\!\!\!\!/\,\, 0} c_n
\Bigl(
e^{-{\rm i}(\omega_n \tau +k_n\sigma)} {\tilde S}^a_n 
- {\rm i}{\textstyle{\omega_n-k_n\over m}}
e^{-{\rm i}(\omega_{n} \tau -k_n\sigma)}
\Pi { S}_n^a
\Bigr)~,
\end{eqnarray}
where we have defined
\begin{equation}
c_n =\frac{1}{\sqrt{1+({\omega_n -k_n\over m})^2}}\ .
\end{equation}
As in the bosonic case, the normalizations are chosen in such a way that
the canonical equal-time anti-commutator 
\begin{equation}
\label{canS}
\{S^a(\sigma),\, S^b(\sigma') \}=
\{{\tilde S}^a(\sigma),\, {\tilde S}^b(\sigma')\}=
\delta^{ab}\delta_{m+n,0}\, \delta(\sigma-\sigma')
\end{equation}
implies for the modes the relations
\begin{equation}
\label{modesacomm}
\{S^a_n,S^b_m\}=
\{{\tilde S}^a_n~,\hskip 0.2cm{\tilde S}^b_m\}=
\delta^{ab}\delta_{m+n,0}~. 
\end{equation}

The vacuum of the fermionic zero modes  sector can be chosen to be
the usual set of states $\{ \ket I\, , \,\ket \ad\}$ which satisfy 
$S^a_0 \ket I =\gamma^I_{a\ad}\, \ket\ad / \sqrt 2$ and
$S^a_0 \ket\ad =\gamma^I_{a\ad}\, \ket I / \sqrt 2$
or, as done in \cite{rrm}, to be the state $\ket 0$ such that 
\begin{equation}
(S^a_0+i\,{\tilde S}^a_0) \ket 0
=0~,
\end{equation}
which is related to the usual vacuum in such a way that it represents the
zero modes part of the flat $D(-1)$ brane.
In the non-zero modes sector the prescription is unique:
\begin{equation}
S^a_n \ket 0 = {\tilde S}^a_n \ket 0~,
\,\,\,\,
n>0
\,\,
.
\end{equation}

\subsection{Constraints on boundary states}
As in \cite{Green:1996um} we look for states which break half of the
supercharges and therefore satisfy
\begin{eqnarray}
\label{QB}
\left( Q_a + i \eta\, M_{(s)\,ab}\, {\tilde Q}_b \right) \ket B&=&0~,
\\
\label{QtB}
\left( Q_\ad + i \eta\, M^{}_{(c)\,\ad\bd}\, {\tilde Q}_\bd \right) \ket B&=&0~.
\end{eqnarray}
These are the same conditions as in \cite{Green:1996um} because they are not
affected by the breaking of the $SO(8)$ symmetry, as
the reader can convince her/himself by decomposing the charges in $SO(4)$
chiral blocks and then reassembling the expressions written with these
blocks. The $SO(8)$ breaking only affects the constraints the matrices $M$ 
have to satisfy, as we will see.

Differently from what happens in flat space, some of the supersymmetry charges
do not commute with the hamiltonian, therefore we must impose explicitely 
that the constraints (\ref{QB},\ref{QtB}) be time invariant.
From the commutation relations
\begin{equation}
[P^-,\, Q_a]= \Pi_{a b} Q_a
\,\,
,
\,\,
[P^-,\, {\tilde Q}_a]= \Pi_{a b} {\tilde Q}_a
\end{equation}
we get a further constraint on the matrix $M_{(s)\,ab}$:
\begin{equation}
\label{new_const}
M_{(s)\,ab}=(\Pi M_{(s)}\Pi)_{\,a b}
\,\,
.
\end{equation}

The supercharges used in the previous expressions are given by \cite{rrm}:
\begin{equation}
\label{q+}
\frac{1}{2^{3/4}\sqrt{ p^+} }Q_{a} =S_0
\,,
\qquad
\frac{1}{2^{3/4} \sqrt{p^+} }\tilde{Q}_{a} = {\tilde S}_0 \,,
\end{equation}
\begin{eqnarray}
\label{q-} 
\frac{\sqrt{ p^+}}{2^{1/4} } Q_{\ad}
&=&  p_0^I\bar{\gamma}^{I\ad b} S_0^b 
   - m x_0^I\left(\bar{\gamma}^I\Pi\right)_{\ad b} 
  {\tilde S}_0^b  
\nonumber\\ 
&+& \sum_{n=1}^\infty
\left( \sqrt{2\omega_n}c_n \bar{\gamma}^{I}_{\ad b} 
       (a_n^{\dagger I} S_n^b+a_n^{ I} S^{b \dagger}_n )
+
 \frac{{\rm i} m}{\sqrt{2\omega_n} c_n }
 \left(\bar{\gamma}^I\Pi\right)_{\ad b} 
       ({\tilde a}_n^{\dagger I} {\tilde S}_n^b
        -{\tilde a}_n^{ I} {\tilde S}^{b \dagger}_n )
\right)~,
\nonumber\\
\\ 
\label{qt-}
\frac{\sqrt{ p^+}}{2^{1/4} } {\tilde Q}_{\ad}
&=& p_0^I\bar{\gamma}^{I}_{\ad b} {\tilde S}_0^b
  + m x_0^I\left(\bar{\gamma}^I\Pi\right)_{\ad b}{ S}_0^b 
\nonumber\\ 
&+& \sum_{n=1}^\infty
\left(( \sqrt{2 \omega_n}c_n  \bar{\gamma}^{I}_{\ad b} 
       ({\tilde a}_n^{\dagger I} {\tilde S}_n^b
        +{\tilde a}_n^{ I} {\tilde S}^{b \dagger}_n )
-
 \frac{{\rm i} m}{\sqrt{2 \omega_n} c_n }
 \left(\bar{\gamma}^I\Pi\right)_{\ad b} 
       (a_n^{\dagger I} S_n^b-a_n^{ I} S^{b \dagger}_n )
\right)~.
\nonumber\\ 
\end{eqnarray}
and the light-cone Hamiltonian is given by
\begin{eqnarray}
\nonumber
P^-=-H_{l.c.}&=&
\frac{1}{2p^+}\left(p_0^2+m^2\, x_0^2
+2i\, S^a_0\, \Pi_{a b} \tilde S^b_0
\right)
\\
&&+\frac{1}{p^+}
\sum_{n=1}^\infty \omega_n 
\left( {a}^\dagger_n {a}_n
      +{S}^{a\dagger}_n {S}^a_n 
      +{\tilde a}^\dagger_n {\tilde a}_n
      +{\tilde S}^{a\dagger}_n {\tilde S}^a_n
\right)~. 
\end{eqnarray}

Multiplying (\ref{QB}) by 
$\left( Q^c + i \eta\, M^{c}_{(s)\,d}\, {\tilde Q}^d \right)$ 
and taking the anticommutator, we get as an immediate consequence 
that
\begin{equation}
\left(M_{(s)}\, M_{(s)}^T\right)_{ab}=\delta_{ab}
\,\, 
.
\end{equation}
In a similar way from (\ref{QtB}), using the anticommutation relations and 
looking at the terms 
proportional to $P^-$ we find
\begin{equation}
\left(M_{(c)}\, M_{(c)}^T\right)_{\ad\bd}=\delta_{\ad\bd}
\,\,
.
\end{equation}
Equation (\ref{QtB}) can be rewritten as
\begin{equation}
\left( S^a_0 + i \eta\, M^a_{(s)\,b}\, {\tilde S}^b_0 \right) \ket B=0~, 
\end{equation}
which suggests to take the following ansatz:
\begin{equation}
\label{ans-S}
\left( S^a_n + i \eta\, M^{a}_{(s)\,b}\, {\tilde S}^b_{-n} \right) \ket B=0~, 
\,\,\,\,
n=0,\pm1,\pm2,\dots~,
\end{equation}
in order to solve the second defining equation (\ref{QtB}).

Using the ansatz (\ref{ans-S}) and (\ref{ans-a}) into (\ref{QtB}), we
get three different equations. From the non zero modes structure
$a^\dagger S+ a S^\dagger$ we find  
\begin{equation}
\label{const1}
M^{JI}\, (\bar\gamma^J)_{\ad b}
=
(M_{(c)} {\bar\gamma}^I M^T_{(s)})_{\ad b}
\,\,
.
\end{equation}
This is the same equation arising in flat space \cite{Green:1996um}.
From the non zero modes structure $a^\dagger S- a S^\dagger$, which always
enters multiplied by $m$, we get a new equation which explicitly breaks
the $SO(8)$ invariance:  
\begin{equation}
\label{const2}
M^{IJ}\, (\bar\gamma^J \Pi)_{\ad b}
=
-(M_{(c)} {\bar\gamma}^I \Pi M_{(s)})_{\ad b}~.
\end{equation}
Finally, the zero-mode sector yields a further constraint which reads
\begin{equation}
\left[ 
p^I \left( {\bar\gamma}^I M_{(s)} 
          -  M_{(c)} {\bar\gamma}^I\right)_{\ad b}
+i\eta\, 
x^I_0 \left( - {\bar\gamma}^I \Pi
           + M_{(c)} {\bar\gamma}^I \Pi M_{(s)} \right)_{\ad b}
\right]\, {\tilde S}^b_0 \ket B =0~,
\end{equation}
which can be  rewritten in a better way with the help of
(\ref{const1},\ref{const2}) as
\begin{equation}
\label{const3}
\left[ {\bar\gamma}^J_{\ad b}\, \left(\delta^{IJ}-M^{IJ}\right) p^I
-i\eta\,  x^I_0 \left(\delta^{IJ}+M^{IJ}\right)\, 
                \left({\bar\gamma}^I \Pi M_{(s)}\right)_{\ad b}
\right]\,{\tilde S}^b_0 \ket B =0~.
\end{equation} 

\subsection{Boundary states.}
Since the constraints (\ref{const1},\ref{const2},\ref{const3}) are
invariant under  $SO(4)\times SO(4)$ rotations as $\Pi$ is, we can
look for special solutions from which we can derive the general ones
by a rotation.
As usual in light-cone formalism  branes are ``instantonic'' since they
have Dirichlet boundary condition on $x^+\propto \tau$.
The complete boundary states are given by
\begin{eqnarray}
\ket B 
&=&
\exp\left( \sum_{n=1}^\infty 
           M_{IJ}\,a_n^{I\dagger} \tilde a_n^{J\dagger}
         -i\eta\, M_{(s)\,ab}\,  S^{a\dagger}_n {\tilde S}^{a\dagger}_n
     \right)
\ket B_0~,
\\
\ket B_0 
& = &
\left( M_{IJ} \ket I \, \ket {\tilde J} 
      +i\eta\, M^{}_{(c)\,\ad\bd} \ket \ad \, \ket {\tilde \bd}
\right)
\,\,
e^{ \frac{1}{2} M_{IJ}\,a_0^{I\dagger} a_0^{J\dagger} } \ket 0_a 
\end{eqnarray}

The explicit form of the matrices $M$ is given, up to $SO(4)\times
SO(4)$ rotations, by the following cases.

\paragraph{D(-1):} 
No solutions preserve 16 supersymmetries.
The natural solution $M_{(v)}=M_{(s)}=M_{(c)}=1_8$ violates (\ref{const2}),
which in turn means that (\ref{QtB}) is violated. Eight supercharges are still
preserved, thanks to (\ref{QB}). 

\paragraph{D1:}
Both spatial Neumann directions must lie in the first four $1,2,3,4$ or second
four directions $5,6,7,8$. Moreover the brane must seat in the origin because
of the zero modes constraints (\ref{const3}).
An explicit solution is given by
\begin{equation}
M^{a}_{(s)\,b}=(\gamma^1\bar\gamma^2)_{ab}~,
\,\,\,
M^{\ad}_{(c)\,\bd}=(\bar\gamma^1 \gamma^2)_{\ad\bd}~,
\,\,\,
M^{II}=
\left\{
\begin{array}{l c}
-1 & I=1,2\\
+1 &I\ne1,2
\end{array}
\right.~,
\,\,\,
q_0^{3,4,5,6,7,8}=0~.
\end{equation}
Relaxing the condition on the zero-modes $q_0$, i.e., allowing the brane to sit
anywhere, violates (\ref{const3}) and thus (\ref{QtB}). Only 8 charges are then
preserved. 
\paragraph{D3:}
No solution preserves 16 supersymmetry charges.
The following would-be solution does not satisfy the constraint 
(\ref{new_const}) for time-invariance.
Three of the four spatial Neumann directions are 
in either the first four $1,2,3,4$ or second four directions
$5,6,7,8$.
Again the brane is fixed at the origin of the transverse Dirichlet coordinates:
$q_0^{4,6,7,8}=0$.
An explicit solution is
\begin{equation}
M^{a}_{(s)\,b}=(\gamma^1 \bar\gamma^2 \gamma^3 \bar\gamma^5)_{ab}~,
\,\,\,
M^{\ad}_{(c)\,\bd}=(\bar\gamma^1 \gamma^2 \bar \gamma^3 \gamma^5)_{\ad\bd}~,
\,\,\,
M^{II}=
\left\{
\begin{array}{l c}
-1 & I=1,2,3,5\\
+1 & I=4,6,7,8
\end{array}
\right.~.
\end{equation}

With an even number of directions in the first or second four directions (and
arbitrary positions $q_0$) one obtains a solution preserving the 8 charges
corresponding to (\ref{QB}).

\paragraph{D5:}
An even number of directions have to be in each of the two sets of
four coordinates. The brane is stuck at the origin of the remaining
coordinates: $q_0^{7,8}=0$ if it must preserve 16 charges (otherwise only 8
are presrved), and
\begin{equation}
M^{a}_{(s)\,b}=(\gamma^1 \bar\gamma^2 \gamma^3 \bar\gamma^4 
\gamma^5 \bar\gamma^6 )_{ab}~,
\,\,\,
M^{\ad}_{(c)\,\bd}=(\bar\gamma^1 \gamma^2 \bar \gamma^3 \gamma^4
\bar\gamma^5 \gamma^6
)_{\ad\bd}~,
\,\,\,
M^{II}=
\left\{
\begin{array}{l c}
-1 & I\ne 7,8\\
+1 & I=7,8
\end{array}
\right.~.
\end{equation}

\paragraph{D7:}
no solutions.\hfill\\
The would-be solution $M_{(v)}=-1_8$,
$M_{(s)}=
\gamma^1 \bar\gamma^2 \gamma^3 \bar\gamma^4 \gamma^5 \bar\gamma^6 \gamma^7$
and
$M_{(c)}=
\bar\gamma^1 \gamma^2 \bar\gamma^3 \gamma^4 \bar\gamma^5 \gamma^6 \bar\gamma^7$
violates (\ref{const2}).

As a further check we can notice that applying $P^I=p^I_0$ to
(\ref{QtB}) we get
\begin{equation}
im\, \left[\left(\bar\gamma^I \Pi 
         - M_{(s)}\,\bar\gamma^I\Pi\,M_{(c)}\right) {\tilde S}_0
    \right]_\ad
+\left( Q_\ad + i \eta\, M^{}_{(c)\,\ad\bd}\, {\tilde Q}_\bd \right)
p^I_0
\ket B =0
\end{equation}
which is trivially satisfied in Neumann directions.

It is nevertheless worth noticing that the B2B amplitude between two
would-be $D(-1)$ boundary states located in arbitrary positions is
{\sl zero}, even if they do not preserve 16 supercharges.
Indeed, the would-be $D(-1)$  boundary state does belong neither 
to a short representation  nor to a long one: 8 charges are still preserved 
since (\ref{QB}) is still satisfied.
Therefore inserting 
$1=\frac{1}{2^{5/2} p^+}\{
Q_a - i \eta\, M_{(s)\,a c}\, {\tilde Q}_c
\, ,
Q_d + i \eta\, M_{(s)\,b d}\, {\tilde Q}_d
\}$
into the boundary-to-boundary amplitude 
$\bra {D(-1),q_0} e^{-\tau\, L_0^{l.c.}} \ket{D(-1),q_1}$ 
we get zero.
More generally the same result is valid for the boundary-to-boundary
 amplitude between
two flat space boundary states whose $M_{(s)}$ do satisfy
(\ref{new_const}): in particular it
 applies to $D3$'s with two directions in each of the two
coordinates sets and to the other branes not at the origin of the
transverse coordinates.
\appendix
\setcounter{section}{0} \setcounter{subsection}{0}
\section{Notation  and definitions}
We essentially use the Metsaev and Tseylin's  conventions \cite{rrm}
which we report here for self consistency.
The conventions for the indices are:
\begin{eqnarray*}
\\
\mu,\nu,\rho = 0,1,\ldots, 9 && \qquad  so(9,1) \  \hbox{ vector
indices (tangent space indices) }
\\
I,J,K,L = 1,\ldots, 8 && \qquad  so(8) \  \hbox{ vector indices
(tangent space indices) }
\\
\alpha,\beta,\gamma = 1,\ldots, 16 && \qquad  so(9,1) \  \hbox{
spinor indices in chiral representation}
\\
a,b,c = 1,\ldots, 8 && \qquad  so(9,1) \  \hbox{
spinor indices surviving the $\kappa$ symmetry} 
\\
&&
\phantom{\qquad  so(9,1) \ }
\hbox{ fixing in chiral representation}
\\
\ad,\bd,\cd = 1,\ldots, 8 && \qquad  so(9,1) \  \hbox{
spinor indices not surviving the $\kappa$ symmetry} 
\\
&&
\phantom{\qquad  so(9,1) \ }
\hbox{ fixing in chiral representation}
\\
A,B = 0,1 && \qquad  \hbox{ 2-d world-sheet coordinate indices}
\end{eqnarray*}
We identify  the transverse target
indices with tangent space indices, i.e.
$x^{\underline{I}} = x^I$,
and avoid  using
the underlined indices in $+$ and $-$ light-cone directions, i.e.
adopt  simplified notation $x^+$, $x^-$. We suppress the flat
space metric tensor $\eta_{\mu\nu}=(-,+,\ldots, +)$ in scalar
products, i.e. $ X^\mu Y^\mu\equiv \eta_{\mu\nu}X^\mu Y^\nu.$
 We decompose
$x^\mu$ into the light-cone and transverse coordinates: $x^\mu=
(x^+,x^-, x^I)$, $x^I=(x^i,x^\ipr)$, where 
\begin{equation}
 x^\pm
\equiv
\frac{1}{\sqrt{2}}(x^9\pm x^0)\,.
\end{equation}
The scalar products of
tangent space vectors are decomposed as 
\begin{equation}
X^\mu Y^\mu = X^+Y^-+ X^-Y^+ +X^IY^I
\,, \qquad 
X^IY^I=X^iY^i+X^\ipr Y^\ipr\,. 
\end{equation}
 The notation $\partial_\pm$, $\partial_I$ is  mostly  used
for target space derivatives\footnote
{In sections 2, 3.2.3 and 4.1
 $\partial_\pm$ indicate world-sheet derivatives.}
\begin{equation}
\partial_+ \equiv \frac{\partial}{\partial x^+}
\qquad
\partial_- \equiv \frac{\partial}{\partial x^-} \,,\qquad
\partial_I \equiv \frac{\partial}{\partial x^I}\,.
\end{equation}
We also use
\begin{equation}
\partial^+ = \partial_-\,,\qquad
\partial^- = \partial_+\,,\qquad
\partial^I = \partial_I\,.
\end{equation}
 The
$SO(9,1)$ Levi-Civita tensor is defined by
$\epsilon^{01\ldots 9}=1$, so that in the light-cone coordinates 
$\epsilon^{+-1\ldots 8}=1$. 
The derivatives with respect to the  world-sheet coordinates
$(\tau,\sigma)$  are  denoted as
\begin{equation}
 \dot{x}^I \equiv \partial_\tau x^I\,, \ \ \ \
 \qquad \x'^I \equiv
\partial_\sigma x^I\,. 
\end{equation}
We use the chiral representation for the $32\times 32$ Dirac
matrices $\Gamma^\mu$ in terms of the $16\times 16 $
matrices $\gamma^\mu$
\begin{equation}
 \Gamma^\mu =\left(\begin{array}{cc} 0  & \gamma^\mu \\
\bar{\gamma}^\mu & 0
\end{array}\right) \,,
\end{equation}
\begin{equation}
\gamma^\mu\bar{\gamma}^\nu + \gamma^\nu\bar{\gamma}^\mu
=2\eta^{\mu\nu}\,,\qquad \gamma^\mu =
(\gamma^\mu)^{\alpha\beta}\,, \qquad  \bar{\gamma}^\mu
=\gamma^\mu_{\alpha\beta}\,, 
\end{equation}
\begin{equation}
\label{gammaspl}\gamma^\mu=(1,\gamma^I,\gamma^9)\,,\qquad
\bar{\gamma}^\mu=(-1,\gamma^I,\gamma^9)\,,\qquad
\alpha,\beta=1,\ldots 16\,.
\end{equation}
We adopt the Majorana
representation for $\Gamma$-matrices, $ C= \Gamma^0$, which
implies that all $\gamma^\mu$ matrices are real and symmetric,
$\gamma^\mu_{\alpha\beta} = \gamma^\mu_{\beta\alpha}$,
$(\gamma^\mu_{\alpha\beta })^* = \gamma^\mu_{\alpha\beta}$. As in
\cite{rrm} $\gamma^{\mu_1\ldots \mu_k}$ are the antisymmetrized
products of $k$ gamma matrices, e.g.,
$(\gamma^{\mu\nu})^\alpha{}_\beta \equiv
\frac{1}{2}(\gamma^\mu\bar{\gamma}^\nu)^\alpha{}_\beta -(\mu
\leftrightarrow \nu)$, $ (\gamma^{\mu\nu\rho})^{\alpha\beta}
\equiv
\frac{1}{6}(\gamma^\mu\bar{\gamma}^\nu\gamma^\rho)^{\alpha\beta}
\pm 5 \hbox{ terms}$. { Note that
$(\gamma^{\mu\nu\rho})^{\alpha\beta}$ are antisymmetric in
$\alpha$, $\beta$.}
 We assume moreover the following block decomposition for $\gamma^I$:
\begin{equation}
\gamma^I 
=\left(\begin{array}{cc}
0 & (\gamma^I)^{a\bd} \\
(\gamma^I)^{\ad b} & 0
\end{array}\right)
=
\left(\begin{array}{cc}
0 & (\tau^I) \\
(\tau^I)^T & 0
\end{array}\right)
\end{equation}
and the normalization
\begin{equation}
\label{g11} 
\Gamma_{11} \equiv  \Gamma^0\ldots \Gamma^9
=\left(\begin{array}{cc}
1_{16} & 0\\
0 & -1_{16}
\end{array}\right)\ 
, \ \ 
 \gamma^0\bar{\gamma}^1 \ldots \gamma^8\bar{\gamma}^9= 1_{16}
,\ 
\gamma^+ ={\bar \gamma}^-
=
\sqrt2
\left(\begin{array}{cc}
1_8 & 0\\
0 & 0_8
\end{array}\right)\
\end{equation}
We use the following definitions
\begin{equation}
\Pi^\alpha{}_\beta \equiv
(\gamma^1\bar{\gamma}^2\gamma^3\bar{\gamma}^4)^\alpha{}_\beta\,,\qquad
 (\Pi^\prime)^\alpha{}_\beta \equiv
(\gamma^5\bar{\gamma}^6\gamma^7\bar{\gamma}^8)^\alpha{}_\beta\,.
\end{equation}
\begin{equation}
\bar{\Pi}_\alpha{}^\beta \equiv
(\bar{\gamma}^1\gamma^2\bar{\gamma}^3\gamma^4)_\alpha{}^\beta\,,\qquad
 (\bar{\Pi}^\prime)_\alpha{}^\beta \equiv
(\bar{\gamma}^5\gamma^6\bar{\gamma}^7\gamma^8)_\alpha{}^\beta\,.
\end{equation}
{Note that
$\Pi^\alpha{}_\beta=\bar{\Pi}_\beta{}^\alpha$}.
Because of the relation $\gamma^0\bar{\gamma}^9 =\gamma^{+-}$ the normalization
condition \ref{g11} takes the form $ \gamma^{+-}\Pi\Pi^\prime =
1$. Note also the following useful relations (see also \cite{rrm})
\begin{equation}
 (\gamma^{+-})^2 = \Pi^2 =(\Pi^\prime)^2 =1\,,
\end{equation}
\COMMENTO{
\begin{equation}
\gamma^{+-}\gamma^{\pm} =\pm \gamma^\pm\,,\qquad \bar{\gamma}^\pm
\gamma^{+-} = \mp \bar{\gamma}^\pm\,,\qquad
\gamma^+\bar{\gamma}^+ = \gamma^-\bar{\gamma}^- =0\,,
\end{equation}
\begin{equation}
\bar{\gamma}^+(\Pi+\Pi^\prime)=(\Pi+\Pi^\prime) \gamma^- =0\,,
\qquad \bar{\gamma}^-(\Pi - \Pi^\prime)= (\Pi - \Pi^\prime)
\gamma^+ =0\,. 
\end{equation}
\begin{equation}
\gamma^\pm \bar{\Pi} = \Pi
\gamma^\pm\,,\quad \gamma^i \bar{\Pi} = -\Pi\gamma^i,\quad
\bar{\gamma}^i\Pi = - \bar{\Pi}\bar{\gamma}^i,\quad
\gamma^i\bar{\Pi}' = \Pi'\gamma^i\,, \quad \bar{\gamma}^i\Pi' =
\bar{\Pi}'\bar{\gamma}^i\,.
\end{equation}
}
\begin{equation}
\tau_I\tau^T_J+\tau_J\tau^T_I
= 2\delta_{IJ}\, 1_8
\end{equation}
 The 32-component positive chirality spinor
$\theta$ and  the negative chirality spinor $Q$  are decomposed
in terms of the 16-component spinors as
\begin{equation}
\theta = \left( \begin{array}{c} 
\theta^\alpha \\
0\end{array}\right)
\,, \qquad\quad
\Gamma^{+-}\theta=
\frac{1}{\sqrt{2\sqrt2} }
\left( \begin{array}{c} 
S^a \\
0_8\\
0_{16}
\end{array}\right)\,
\,, \qquad\quad
Q = 
\left( \begin{array}{c} 
0 \\
Q_\alpha
\end{array}
\right)\,.
\end{equation}
\COMMENTO{
The   complex
Weyl spinor $\theta$ is related to the  two real
Majorana-Weyl spinors $\theta^1$ and $\theta^2$  by
\begin{equation}
\label{comrea} 
\theta = \frac{1}{\sqrt{2}}(\theta^1 +{\rm i}\theta^2)
\,,\qquad 
\bar{\theta} = \frac{1}{\sqrt{2}}(\theta^1 -
{\rm i}\theta^2)\,.
\end{equation}
The short-hand notation like
$\bar{\theta}\bar{\gamma}^\mu\theta$ and $\bar{\gamma}^\mu\theta$
 stand for
$\bar{\theta}{}^\alpha\gamma_{\alpha\beta}^\mu\theta^\beta$ and
$\gamma_{\alpha\beta}^\mu \theta^\beta$ respectively.
}


\end{document}